\documentclass[aps,prb,twocolumn,showpacs,groupedaddress]{revtex4-1}
\usepackage{graphicx}  
\usepackage{dcolumn}   
\usepackage{bm}        
\usepackage{amssymb}   
\usepackage{amsmath}
\usepackage{braket}
\usepackage[mathscr]{euscript}
\usepackage{color}
\begin{document}
\title{Adatoms in graphene nanoribbons: conductance characteristics and quantum spin Hall effect}
\author{Sudin Ganguly}
\email{sudin@iitg.ernet.in}
\author{Saurabh Basu}
\email{saurabh@iitg.ernet.in}
\affiliation{Department of Physics, Indian Institute of Technology Guwahati\\ Guwahati-781039, Assam, India}
\date{\today}
\begin{abstract}
We study the charge and spin transport in a two terminal graphene nanoribbon (GNR) decorated with random Gold (Au) adatoms using a Kane-Mele model. Two commonly used GNRs, that is, the armchair graphene nanoribbon (AGNR) and the zigzag graphene nanoribbon (ZGNR) are compared and contrasted, which shows that in presence of Au adatoms, a somewhat robust $2e^2/h$ conductance plateau occurs in the case of AGNR around the zero of the Fermi energy, while in ZGNR this plateau is fragile. We show that this flat plateau, having a conductance value $2e^2/h$, is not a hallmark signature of a topologically nontrivial quantum spin Hall (QSH) state. Further the conductance decreases by a small amount with the density of adatoms. On the other hand, the spin polarized conductance shows distinct features of enhanced conductivity with increasing Au adatom concentration. Further the fluctuations of the spin polarized conductance have features that carry striking resemblance with the charge conductance profile of the GNRs.
\end{abstract}
\maketitle
%%
%%
%%-----------------------------------INTRODUCTION----------------------------------------------
%%
%%
\section{\label{sec1}Introduction}
After the successful fabrication of graphene \cite{novo}, it has attracted a wide attention in both experimental and theoretical investigations. Graphene has a linear dispersion near the Dirac points \cite{wallace} which leads to several interesting transport properties \cite{neto}, such as unconventional quantum Hall effect \cite{novo,zhang,vp}, half metallicity \cite {jun,lin} and high carrier mobility \cite{du,bolotin}. These features make graphene a promising candidate for applications in nanoelectronic and spintronic devices. On a parallel front, Kane and Mele \cite{kane-mele1,kane-mele2} predicted that quantum spin Hall (QSH) state can be observed in presence of intrinsic spin-orbit coupling (SOC) which can be created by a complex next nearest hopping that differentiates a left hop from a right one, which triggered an enormous study on topologically nontrivial electronic materials \cite{bernevig,moore,hasan,qi}. However, the QSH effect in clean graphene is still not observed experimentally owing to its vanishingly small intrinsic spin-orbit coupling (SOC) strength \cite{min,yao}, whereas in strong SOC materials, such as CdTe/HgTe quantum wells, the QSH effect has been observed \cite{konig}.

It was theoretically proposed that adsorption of adatoms such as Indium (In), Thallium (Tl), Gold (Au) etc. can enhance or induce intrinsic SOC or Rashba SOC in graphene \cite{weeks,hu,jiang}. While the intrinsic SOC required for the predicted QSH effect, Rashba SOC is detrimental to it. Adatoms like In or Tl can open up a significant topologically nontrivial gap, different theoretical studies have confirmed that the two systems are indeed stable topological insulators \cite{weeks,jiang, waintal}. On the other hand, Au-like adatoms induce Rashba SOC which dominates over the intrinsic SOC and QSH effect will likely to lose in the competition. Therefore it will be relevant to study the transport properties in Au adatom decorated graphene to illustrate the validity of the above scenario. Moreover, owing to the presence of Rashba SOC induced by Au adatoms features of the spin polarized conductance may provide additional clues on the topological phase. 

Metal atoms adsorbed onto graphene sheets also represent a new way for the development of new electronic or spintronic devices. The electronic, structural, and magnetic properties of transition metals on graphene sheets \cite{chan,av,dw} and graphene nanoribbons (GNR) \cite{kan,sevin,rigo,brito} have been studied extensively, which are mostly based on ab-initio density-functional theory (DFT). The spin dependent transport in GNR in presence of Rashba SOC has been investigated in some cases, such as spin filtering effect in zigzag GNR \cite{liu}, possible spin polarization directions for GNR with Rashba SOC \cite{chico}, effects of spatial symmetry of GNR on spin polarized transport \cite{qzhang} etc. However, there are very few studies on the spin dependent transport of Au decorated adatoms that discuss the spin Hall effect and features like nonlocal resistance etc \cite{tuan}. For a recent review, see Ref \cite{roche}.

It is well known that the electronic properties of GNRs depend on the geometry of the edges and lateral width of the nanoribbons \cite{nakada}, and according to the edge termination type, mainly there are two kinds of GNR, namely armchair graphene nanoribbon (AGNR)and zigzag graphene nanoribbon (ZGNR). The ZGNRs are always metallic with zero band gap, while the AGNRs are metallic when the lateral width $N = 3M-1$ ($M$ is an integer), else the AGNRs are semiconducting in nature \cite {fujita} with a finite band gap. 

Since Rashba SOC is the key factor for the spin polarized transport which arises owing to the lack of surface inversion symmetry in a system, in this paper our aim will be on the exploration of spin dependent transport properties of Au decorated adatoms in both kinds of GNR, namely the ZGNR and the AGNR and comment on the presence of the QSH phase therein.

We organize our paper as follows. In the following section, we present for completeness, the theoretical formalism leading to the expressions for the charge and spin polarized conductances using the well known Landauer-B\"{u}ttiker formula. After that we include an elaborate discussion of the results. Where, we have tried to resolve few queries, such as whether the QSH state exists in Au decorated adatoms, how the spin polarized conductance behave in the above system and so on. We have also included an interesting comparison for the conductance properties for the case of AGNR vis-a-vis that of ZGNR.
%%
%%END OF INTRODUCTION
%%
%%---------------------------------Theroy----------------------------------------
%%
%%
\section{\label{sec2}Theoretical formulation and model}

\begin{figure}[h]
\begin{center}
\includegraphics[width=0.45\textwidth,height=0.35\textwidth]{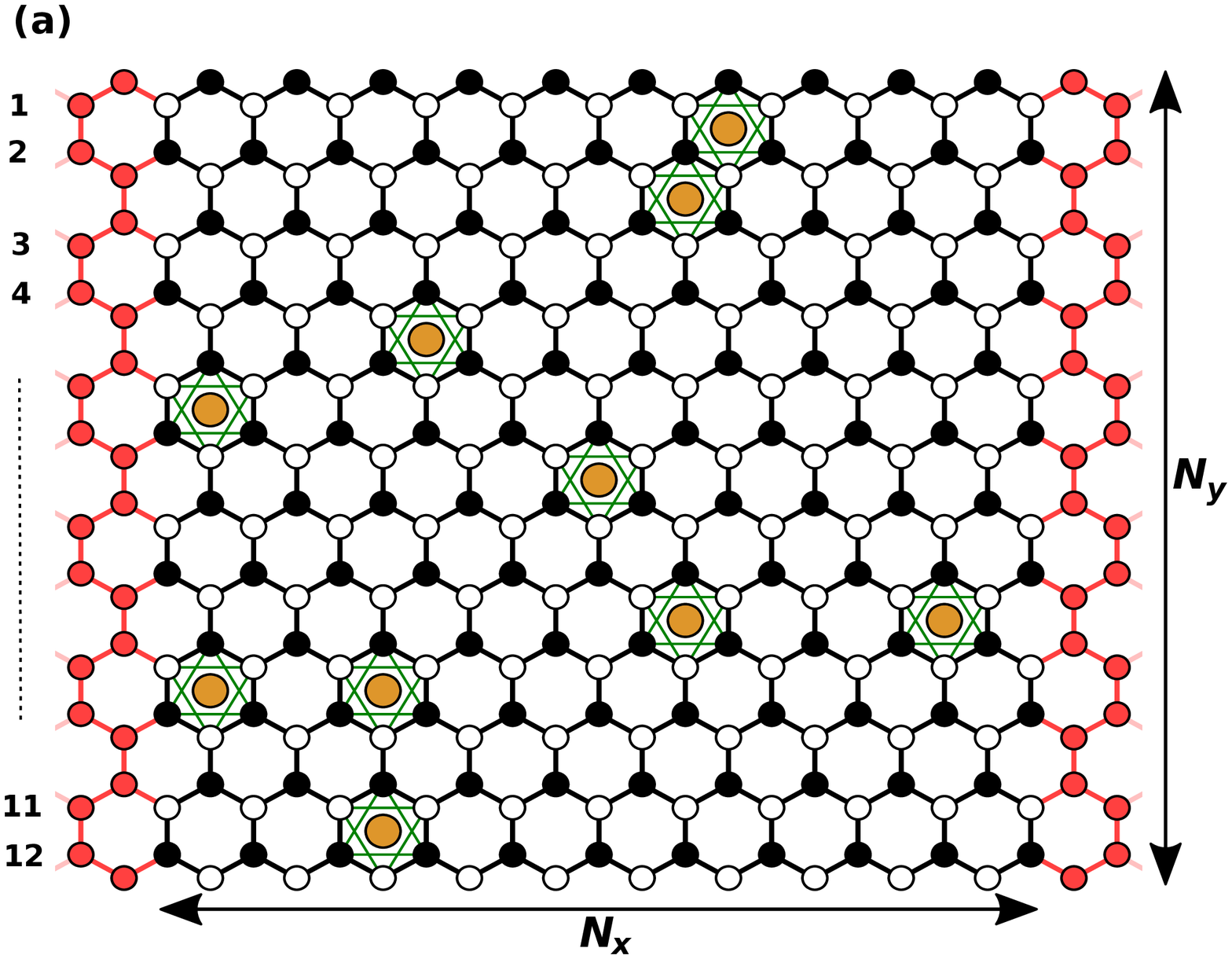}\vskip 0.1 in\includegraphics[width=0.45\textwidth,height=0.35\textwidth]{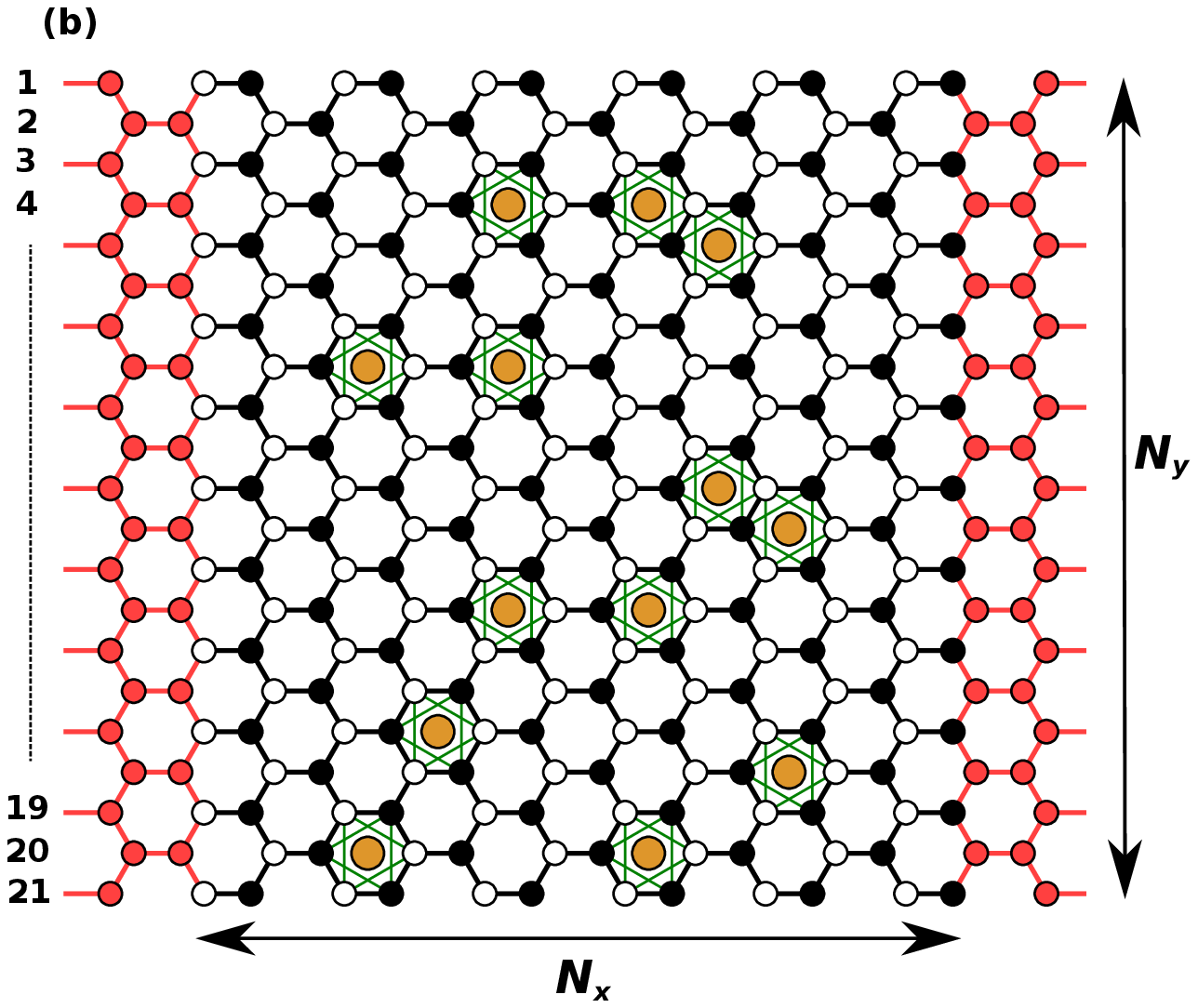}
\caption{Schematic view of a two terminal graphene nanoribbon. (a) Zigzag nanoribbon and (b) armchair nanoribbon. The black and white circles represent the A and B sublattices of graphene. The golden circles are the Au adatoms. The green line is for the next nearest neighbour hopping, which represents the intrinsic SOC, while the black lines surrounding the Au atoms correspond to nearest neighbour hopping and Rashba SOC. Rest of the black lines contain only nearest neighbour hopping. The leads are attached at both ends, which are denoted by red color and are semi-infinite in nature. The leads are free of any kind of SOC. $N_x$ and $N_y$ are the length and width of the nanoribbon respectively.}
\label{setup}
\end{center}
\end{figure}

To begin with we describe the geometry of the system and make our notations clear. We consider a graphene sheet adsorbed with Au atoms, which prefer to reside at the center of the carbon rings \cite{weeks} where it can interact only with the surrounding six-nearest carbon atoms and can enhance the intrinsic SOC or induce Rashba SOC. The effective tight-binding Hamiltonian for graphene with such adatoms is given by \cite{weeks,waintal,kane-mele1},
\begin{eqnarray}
H= - 
t\sum\limits_{\langle ij\rangle}c_i^{\dagger} c_j + 
i\lambda_{SO}\sum\limits_{\langle \langle ij \rangle\rangle\in\mathcal{R}} \nu_{ij} c_i^{\dagger}s^z c_j \nonumber \\ +
i\lambda_R\sum\limits_{\langle ij\rangle\in\mathcal{R}}c_i^{\dagger} \left( {\bf s} \times {\bf\hat{d}}_{ij}\right)_z c_j -  
\mu\sum\limits_{i\in \mathcal{R}} c_i^{\dagger}c_i
\label{h2}
\end{eqnarray} 
where $c_i^{\dagger}=\left(c_{i\uparrow}^{\dagger} \quad c_{i\downarrow}^{\dagger}\right)$ is the creation operator of electrons at site $i$. The first term is the nearest neighbour hopping term, with a hopping strength, $t=2.7$ eV. The second term is the local intrinsic spin-orbit coupling term enhanced by the adatoms residing on the set of hexagons $\mathcal{R}$ that are inhabited by the Au adatoms. $\lambda_{SO}$ is the strength of the intrinsic SOC. The third term is the nearest neighbour Rashba term which explicitly violates $z\rightarrow -z$ symmetry. The last term is the on-site potential, $\mu$ of the carbon atoms in the hexagons hosting adatoms, which describe the chemical potential that screens charge from the adatoms.

The zero temperature conductance $G$ that
denotes the charge transport measurements, is related with
the transmission coefficient as in \cite {land_cond,land_cond2},
\begin{equation}
G = \frac{e^2}{h} T(E)
\end{equation}

The Transmission coefficient can be calculated via \cite{caroli,Fisher-Lee},

\begin{equation}
T = \text{Tr}\left[\Gamma_R {\cal G}_R
  \Gamma_L {\cal G}_A\right]
\end{equation}
${\cal G}_{R(A)}$ is the retarded (advance) Green's function. $\Gamma_{L(R)}$ are the coupling matrices representing the
coupling between the central region and the left (right) lead. They are
defined by the relation \cite{dutta},
\begin{equation}
\Gamma_{L(R)} = i\left[\Sigma_{L(R)} -
  (\Sigma_{L(R)})^\dagger\right]
\end{equation}
Here $\Sigma_{L(R)}$ is the retarded self-energy associated with the left (right) lead. The self-energy contribution is computed
by modeling each terminal as a semi-infinite perfect wire \cite{nico}.

Also the spin polarized conductance can be calculated from \cite{chang},
\begin{equation}
G^s_\alpha = \frac{e^2}{h} \text{Tr}\left[\hat{\sigma}_\alpha\Gamma_R G_R
  \Gamma_L G_A\right]
\end{equation}
Where, $\alpha=x,y,z$ and $\sigma$ denote the Pauli matrices.
Further we define the fluctuations in the spin polarized conductance as,
\begin{equation}
\Delta G^s_\alpha = \sqrt{\langle \left(G^s_\alpha\right)^2\rangle - \langle G^s_\alpha\rangle^2}
\label{fl}
\end{equation}
where  $\langle ... \rangle$ denotes averaging being done over an ensemble of samples with different distributions of adatoms for a particular adatom density $n_{ad}$.

Fig.\ref{setup} shows the geometry used for the calculations of charge and spin polarized conductances. Fig.\ref{setup}(a) is the setup corresponds to ZGNR while Fig.\ref{setup}(b) represents that of AGNR. The length and width of these systems can be determined as shown in the given figure. The systems for example, in Fig.\ref{setup}(a), the width is, $N_y = 12$ and the length is, $N_x = 21$. Thus we can denote the zigzag setup by $N_x$Z-$N_y$A $=$ 21Z-12A. Likewise we may denote the the armchair setup by 12A-21Z (see Fig.\ref{setup}(b)).

The black and white circles stand for the A and B sublattices of graphene. The golden circles are the Au adatoms. The green line is for the next nearest neighbour hopping, which represents the intrinsic SOC, while the black lines surrounding the Au adatoms correspond to the nearest neighbour hopping and the Rashba SOC. Rest of the black lines denote only nearest neighbour hopping. The leads are semi-infinite in nature, attached at both ends and are denoted by red color. The leads are considered to describes by a pure tight binding graphene lattice and hence are free of any kind of SOC.
%%
%%
%%Theory ENDS HERE
%%
%%----------------------------Results-----------------------------------------
%%
%%
\section{\label{sec3}Results and Discussions}

We have investigated the effect of the conducting edges of AGNR and ZGNR in presence of intrinsic and Rashba SOC induced by the adatoms on the experimentally measurable quantity, namely the two terminal charge conductance ($G$) and spin polarized conductance $\left(G^s_\alpha,\quad \alpha=x,y,z \right)$. We have also studied the effect of the orientation of the quantization axis of spin on the spin Hall conductance.

Before embarking on the results, we briefly describe the values of different parameters used in our calculation. Throughout our work, we take the ZGNR setup as 89Z-48A and the AGNR setup as 48A-89Z (see Fig.\ref{setup}). We set the hopping term, $t=2.7$ eV. All the energies are measured in unit of $t$. The charge and spin polarized conductance are measured in units of $\frac{e^2}{\hbar}$. Also the lattice constant, $a$ is taken to be unity. All the measurable quantities are averaged over 100 independent random-adatom configurations for different adatom concentrations, $n_{ad}$. In this work, we have considered three different Au adatom concentrations, namely, $n_{ad} = 0.1,0.2$ and $0.3$. For most of our numerical calculations we have used KWANT \cite{kwant}.

\begin{figure}[h]
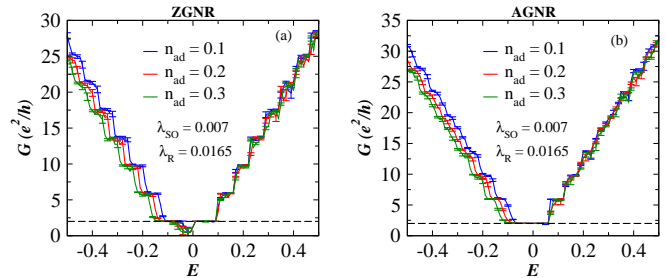

\begin{center}
\includegraphics[width=0.23\textwidth]{fig2a.eps}\hfill\includegraphics[width=0.23\textwidth]{fig2b.eps}
\caption{(Color online) $G$ is plotted as a function of Fermi energy in case of Au adatoms, where $\lambda_{SO}=0.007$, $\lambda_R = 0.0165$ and $\mu = 0.1$ for (a) ZGNR case and (b) AGNR case. The $2e^2/h$ plateau is missing in the ZGNR setup but is present in the AGNR case.}
\label{au_cond}
\end{center}
\end{figure}

The signature for the topological insulator or the QSH phase is that there exists a $2e^2/h$ conductance plateau and the system conducts via the edge states only, while the system in bulk remains insulating in nature. Fig.\ref{au_cond} shows the variation of the conductance as a function of the Fermi energy, $E$ with Au adatoms for three different adatom densities, namely $n_{ad}=0.1,0.2$ and $0.3$. From the first-principles calculations \cite{weeks,tuan}, we use the following parameters: $\lambda_{SO} = 0.007$, $\lambda_R = 0.0165$ and $\mu = 0.1$ (all in units of hopping $t$). The dotted black line corresponds to $2e^2/h$ conductance. 

\begin{figure}[h]
\begin{center}
\includegraphics[width=0.23\textwidth,height=0.2\textwidth]{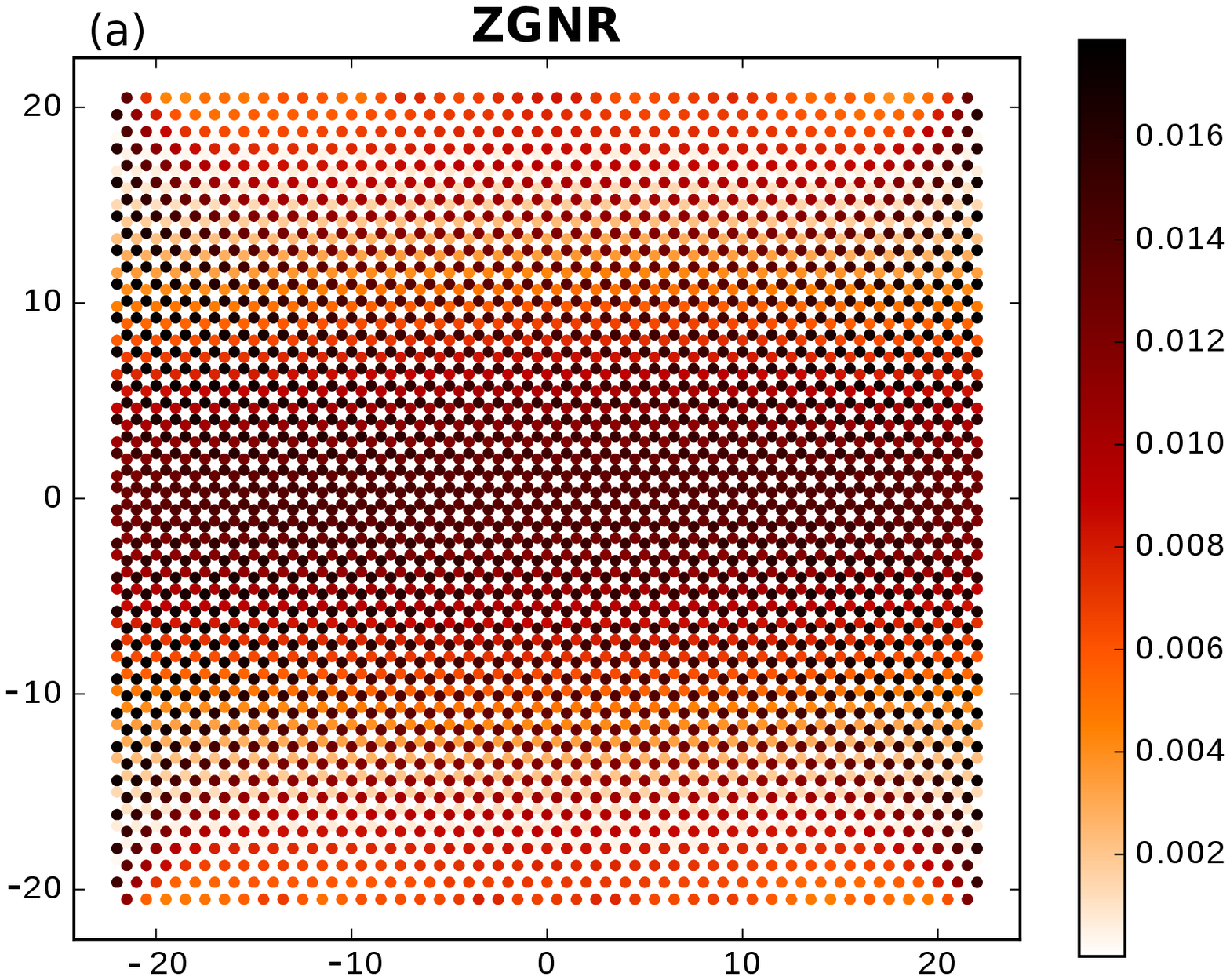}\hfill\includegraphics[width=0.23\textwidth,height=0.2\textwidth]{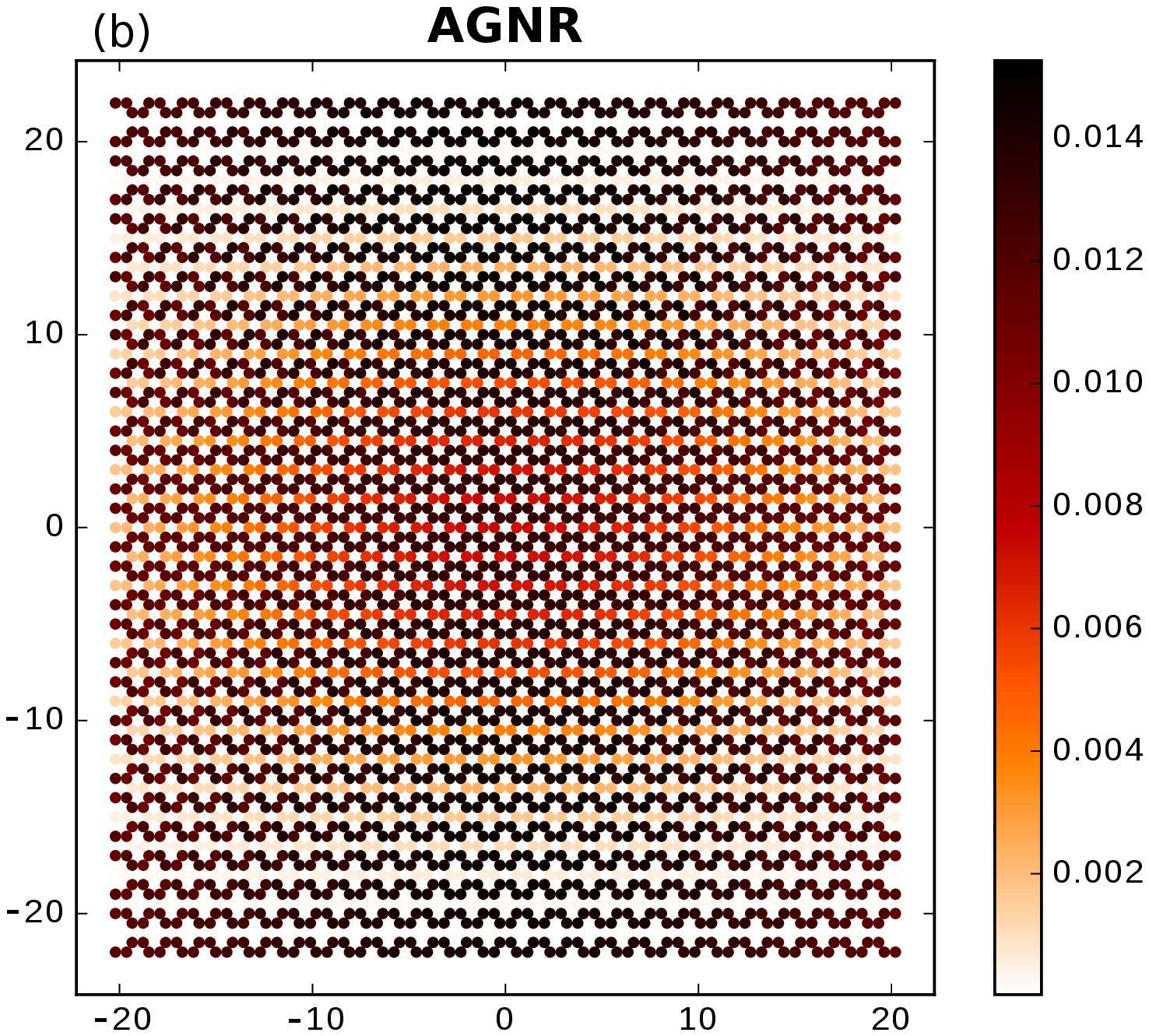}
\caption{(Color online) LDOS plot for (a) ZGNR and (b) AGNR in case of Au adatoms. Adatom concentration is taken to $n_{ad}=0.3$. In both the figures, the system is conducting on the whole. Thus the $2e^2/h$ plateau does not guarantee QSH states.}
\label{au_ldos}
\end{center}
\end{figure}

For the ZGNR setup as shown in Fig.\ref{au_cond}(a), around the zero of the Fermi energy, there is a dip in the conductance to value close to zero. On the other hand, for the AGNR setup (Fig.\ref{au_cond}(b)), a $2e^2/h$ conductance plateau occurs around the zero of the Fermi energy. This leads to a straight forward question, that in case of AGNR setup, can the Rashba SOC (which is a dominant contribution here) stabilize the QSH phase. To answer that question, we plot the space resolved density of states (LDOS) for the AGNR setup as shown in Fig.\ref{au_ldos}(b), where the bulk states yield non-vanishing contributions and hence are conducting. As said earlier, it confirms that Rashba SOC is detrimental for observing the QSH effect, which reveals in our results in case of Au adatom. 

Thus the $2e^2/h$ plateau in the conductance behaviour does not guarantee the presence of a nontrivial topological state \cite{jelena}. This is the one of the interesting results of this work.

\begin{figure}[h]
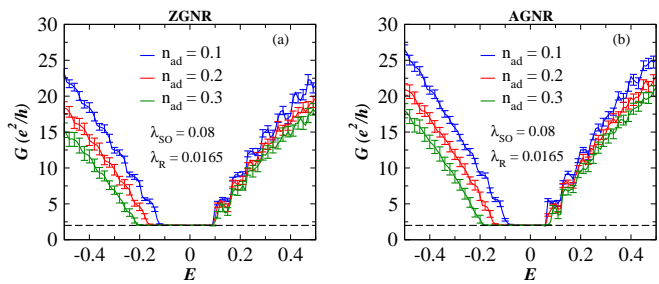

\begin{center}
\includegraphics[width=0.23\textwidth]{fig4a.eps}\hfill\includegraphics[width=0.23\textwidth]{fig4b.eps}
\caption{(Color online) Conductance $G$ is plotted as a function of Fermi energy with $\lambda_{SO}=0.08$ for (a) ZGNR and (b) AGNR. Rest of the parameters are same as in case of Au adatom. Again a $2e^2/h$ conductance plateau occurs in both the cases.}
\label{au_lso0.08}
\end{center}
\end{figure}

\begin{figure}[h]
\begin{center}
\includegraphics[width=0.23\textwidth,height=0.2\textwidth]{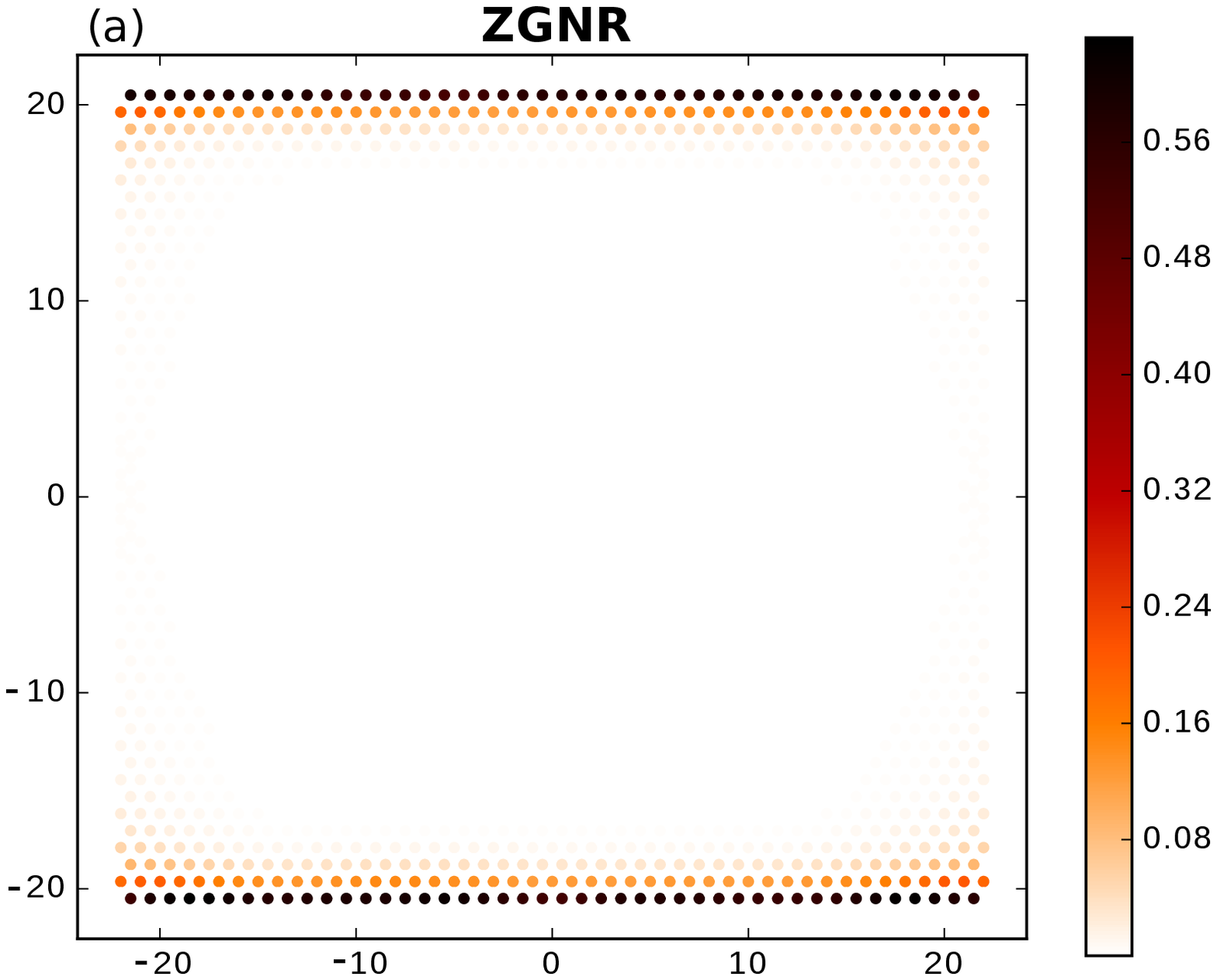}\hfill\includegraphics[width=0.23\textwidth,height=0.2\textwidth]{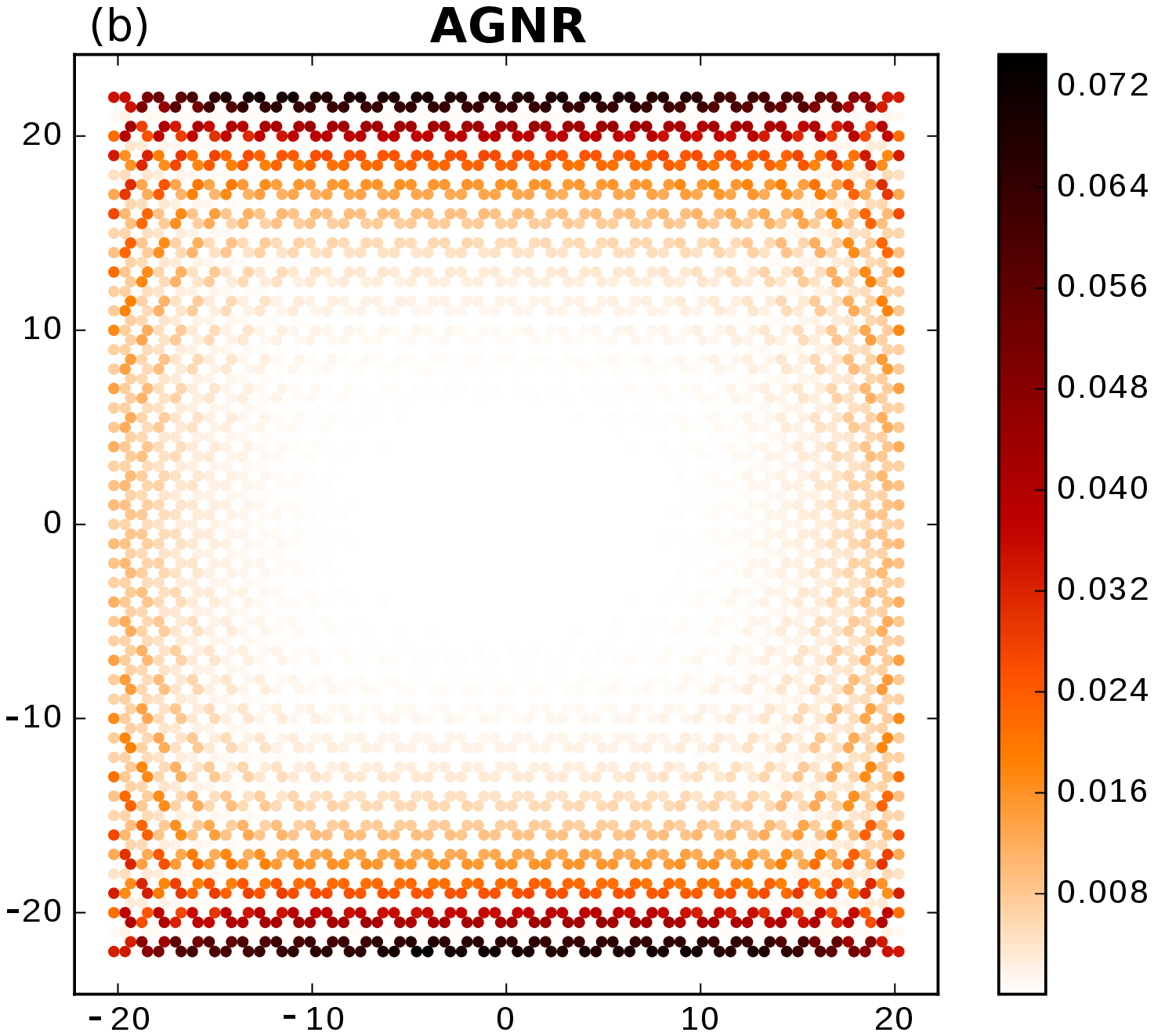}
\caption{(Color online) LDOS plots for (a) ZGNR and (b) AGNR case with $\lambda_{SO}=0.08$ and other parameters same as that of Au adatom for adatom density, $n_{ad}=0.3$. Edge states are again conducting with an insulating nature for the bulk states. Thus by increasing the strength of the intrinsic SOC, we recover the QSH effect.}
\label{au_lso0.08_ldos}
\end{center}
\end{figure}

Hence it boils down to the fact that by some means if we are able to enhance the intrinsic SOC by one order of magnitude compared to the value present in the Au decorated graphene, there could be a way to restore the QSH effect. In Fig.\ref{au_lso0.08}, we show the variation of conductance as a function of Fermi energy for the following parameters: $\lambda_{SO} = 0.08$, $\lambda_R = 0.0165$ and $\mu = 0.1$, that is all the other parameters are same as in case of Au adatom, except the intrinsic SOC. Clearly the $2e^2/h$ conductance plateau re-emerges for the ZGNR setup (Fig\ref{au_lso0.08}(a)) as well as in case of AGNR (Fig\ref{au_lso0.08}(b)). The LDOS plots also support our claim as shown in Fig\ref{au_lso0.08_ldos}(a) and Fig\ref{au_lso0.08_ldos}(b). We also note that there is a certain threshold value for the intrinsic SOC to recover the QSH effect. The threshold is independent of the system size, at least for the few system dimensions checked by us. Specifically, it is observed that when the strength of the intrinsic SOC has the same order of magnitude as that of the chemical potential, the bulk states become insulating in nature and the edge states start to conduct.

Without applying an external magnetic field, spin polarized conductance can be achieved if a Rashba SOC is present in a system. In pristine graphene, though the strength of Rashba SOC is very weak, still it can cause measurable spin polarized conductance. It has been shown that, the $y$-component of spin polarized conductance, $G^s_y$ can be achieved in graphene nanoribbons\cite{chico}. Zhang {\it et al.} \cite{qzhang,qzhang2} showed that the $x$ and $z$ component of the spin polarized conductances $\left(G^s_x,G^s_z\right)$ are zero for ideal graphene nanoribbon because of the longitudinal mirror symmetry of an infinite system \cite{chico}. Adsorption of Au adatoms onto GNR can break this longitudinal mirror symmetry and one can expect non-zero values of $G^s_x$ and $G^s_z$. Motivated by this, we study the behaviour of spin polarized conductance in presence of Au adatoms.

\begin{figure}[h]
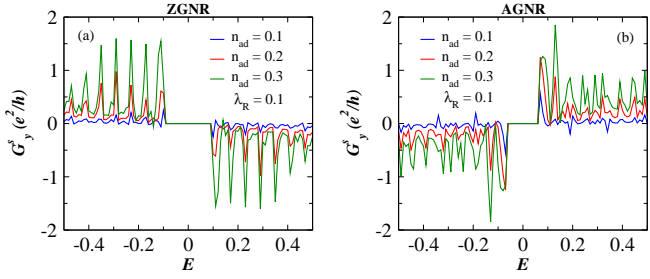

\begin{center}
\includegraphics[width=0.23\textwidth]{fig6a.eps}\quad\includegraphics[width=0.23\textwidth]{fig6b.eps}
\caption{(Color online) The $y$-component of the spin polarized conductance, $G^s_y$ is plotted as a function of the Fermi energy for $\lambda_R=0.1$ for (a) ZGNR and (b) AGNR). Magnitude of $G^s_y$ increases with increasing adatom density.}
\label{sp_lr}
\end{center}
\end{figure}

We begin our study for the spin polarized conductance by taking a hypothetical adatom which induces Rashba SOC among the neighbouring carbon atoms surrounding by the adatoms. Specifically we take the strength of the Rashba SOC, $\lambda_R = 0.1$ which is one order of magnitude large than that of Au adatoms. Fig.\ref{sp_lr} shows the variation of the $y$-component of the spin polarized conductance $\left(G^s_y\right)$ as a function of the Fermi energy for three different adatom concentrations, namely $n_{ad} = 0.1,0.2$ and $0.3$. Fig.\ref{sp_lr}(a) is the result for the ZGNR case and Fig.\ref{sp_lr}(b) for the AGNR. 

We find a few interesting features in the variation of $G^s_y$. The spin polarized conductance is anti-symmetric about $E=0$, which is owing to the electron-hole symmetry \cite{chico,moca}. Also for $E<0$, $G^s_y$ is positive (negative) for the ZGNR (AGNR) setup. The AGNR setup is basically the $\pi/2$ rotated version of the ZGNR setup about the $z$-axis. Under a $\pi/2$ rotation about $z$-axis, the Hamiltonian for the Rashba SOC changes its sign. Specifically, if we define a Rashba Hamiltonian $H_R$ by, $H_R = \sigma_xp_y - \sigma_yp_x$, then $H_R \xrightarrow{{\pi/2\;\text{rotation}}} -H_R$. Which explains the sign difference between the ZGNR and the AGNR setup. Clearly, the magnitude of the $G^s_y$ increases as we increase the adatom density. The spike like features originate due to the finite number of modes available in the leads.

\begin{figure}[h]
\begin{center}
\includegraphics[width=0.23\textwidth]{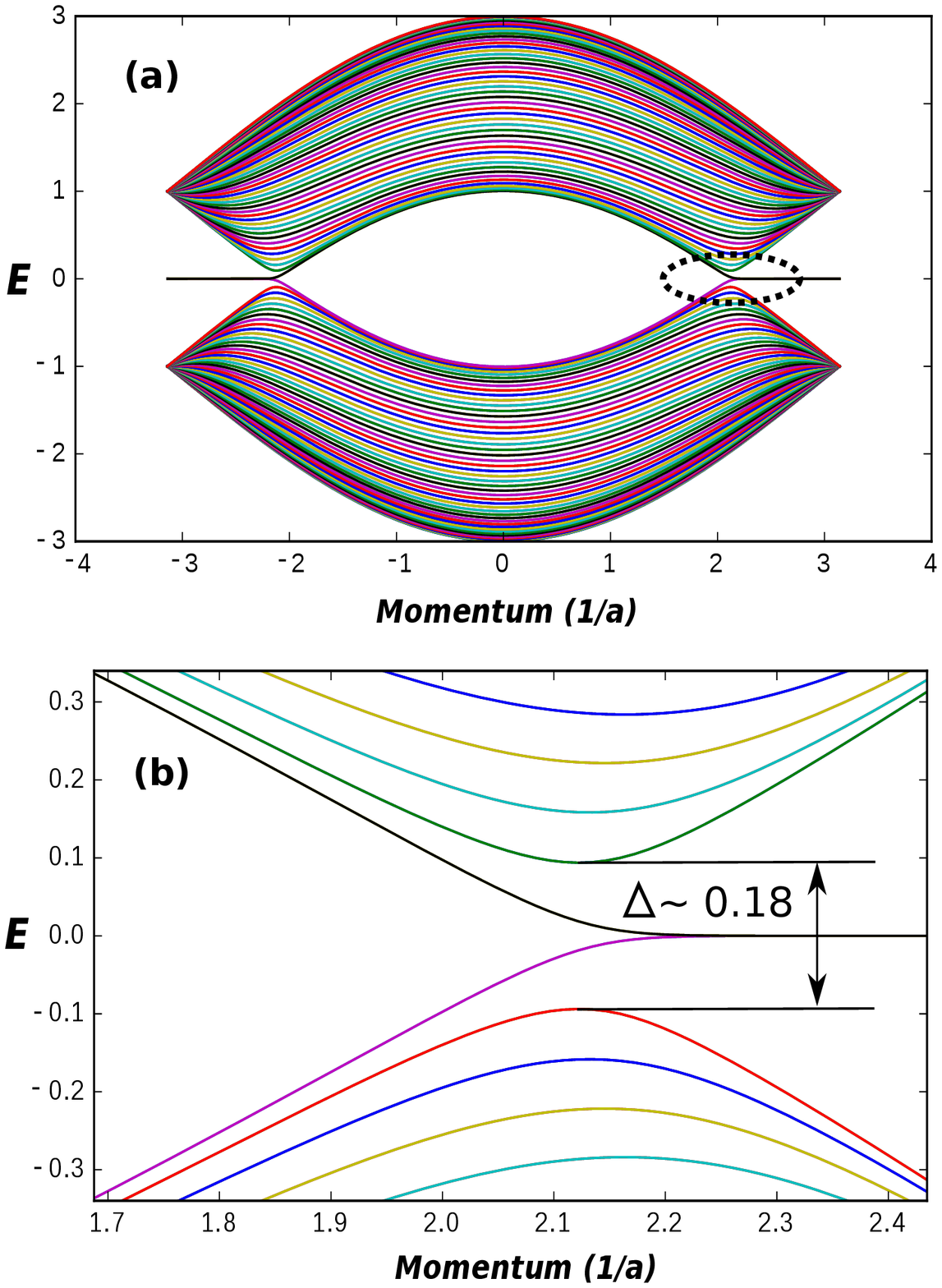}\quad\includegraphics[width=0.23\textwidth]{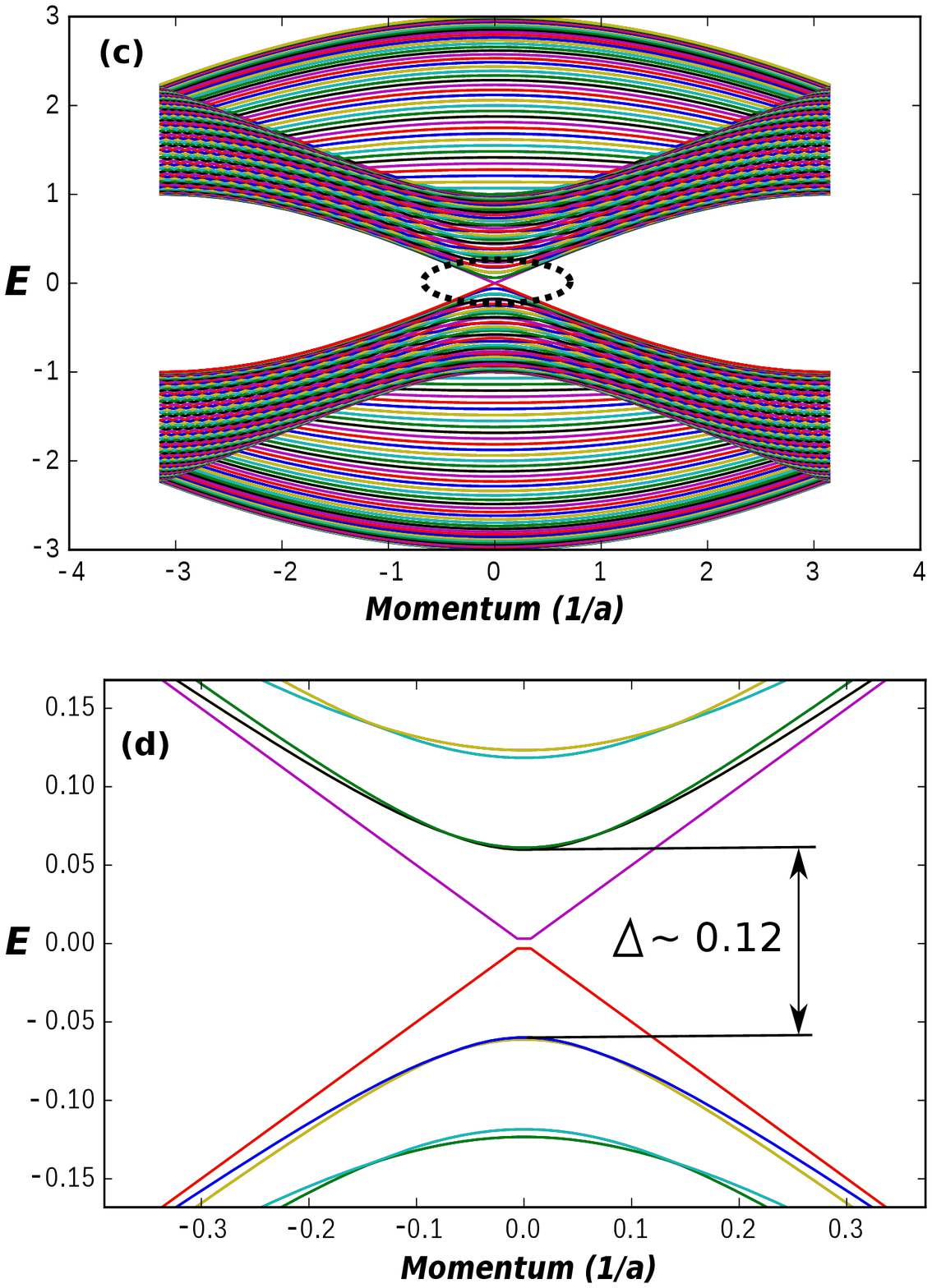}
\caption{(Color online) (a) Band structure of the ZGNR lead. (b) a closer view of ZGNR band structure denoted by the dotted ellipse. (c) and (d) are the band structure for the AGNR case. (b) and (d) show the single channel transmission.}
\label{band}
\end{center}
\end{figure}

Another important point to be noted here, that there is a finite region about the zero energy, where the spin polarized conductance is strictly zero. For the ZGNR case, the the width of the region (in units of $t$) is $\Delta = 0.18$ and $\Delta=0.12$ for the AGNR case. This can be understood from the band structures of the leads corresponding to the ZGNR and AGNR setup as shown in Fig.\ref{band}. Fig.\ref{band}(a) is the band structure for the ZGNR lead and Fig.\ref{band}(b) is a closer view of the region denoted by the dotted ellipse in Fig.\ref{band}(a). Plots in the right panel are for the AGNR lead. There are only two bands in the energy spectrum within the interval, $\Delta$. In other words, this interval corresponds to single channel transmission. Spin polarization is not possible if there is a single channel transmission \cite{chico,liu}. Also this single channel transmission corresponds to the $2e^2/h$ plateau as shown in Fig.\ref{au_lso0.08}.

\begin{figure}[h]
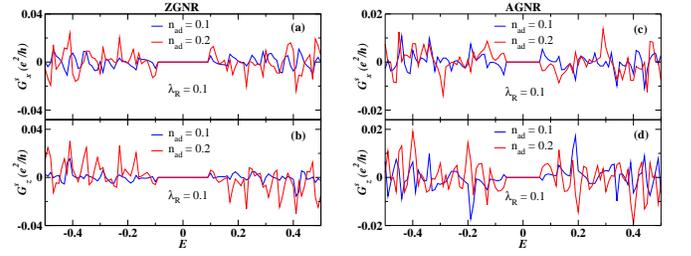

\begin{center}
\includegraphics[width=0.23\textwidth]{fig8a.eps}\hfill\includegraphics[width=0.23\textwidth]{fig8b.eps}
\caption{(Color online) $x$ and $z$-component of the spin polarization are plotted as a function of the Fermi energy for (a) and (b) ZGNR case and (c) and (d) AGNR case with $\lambda_R=0.1$. Only two different adatom density, namely $n_{ad}=0.1,0.2$ are shown for clarity of presentation.}
\label{gxzlr0.1}
\end{center}
\end{figure}

As we introduce the random distribution of adatoms in the graphene nanoribbon, we effectively break the longitudinal mirror symmetry in the system. So it is expected that we should have a non-zero $G^s_x$ and $G^s_z$. Fig.\ref{gxzlr0.1} shows the variation of $G^s_x$ and $G^s_z$ as a function of the Fermi energy for two different adatom densities, namely $n_{ad} = 0.1,0.2$. Fig.\ref{gxzlr0.1}(a) stands for the ZGNR setup while Fig.\ref{gxzlr0.1}(b) for the AGNR. Certainly by destroying the longitudinal mirror symmetry we are able to generate a non-zero $G^s_x$ and $G^s_z$. However, the order of magnitude for the $x$ and $z$-components are two orders of magnitude smaller than the $y$-component of the spin polarized conductance. 

\begin{figure}[h]
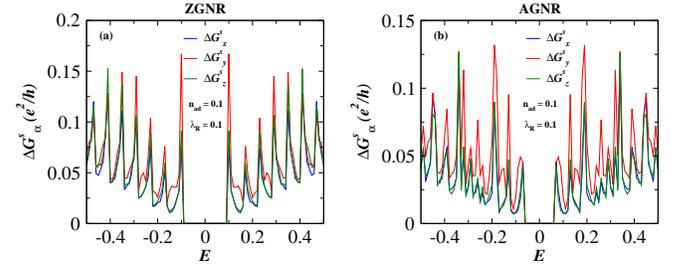

\begin{center}
\includegraphics[width=0.23\textwidth]{fig9a.eps}\quad\includegraphics[width=0.23\textwidth]{fig9b.eps}
\caption{(Color online) The fluctuations in the spin polarized conductance, $G^s_\alpha$ ($\alpha=x,y,z$) is plotted as a function of the Fermi energy for $n_{ad}=0.1$ and $\lambda_R=0.1$ in case of (a) ZGNR  and (b) AGNR. All the three components show similar behaviour.}
\label{dg}
\end{center}
\end{figure}

\begin{figure}[h]
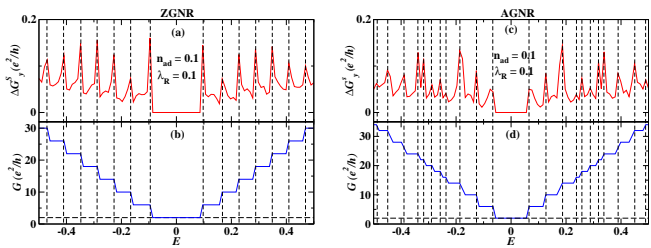

\begin{center}
\includegraphics[width=0.23\textwidth]{fig10a.eps}\quad\includegraphics[width=0.23\textwidth]{fig10b.eps}
\caption{(Color online) $G^s_y$ is plotted as a function of the Fermi energy for (a) ZGNR case and (c) AGNR case with $\lambda_R=0.1$ and adatom density, $n_{ad}=0.1$.  Charge conductance, $G$ is plotted as function of $E$ in the clean limit for (b) ZGNR case and for (d) AGNR case. Vertical dotted lines are shown for comparison.}
\label{spike}
\end{center}
\end{figure}

Since we are considering an average over 100 random configurations in order to calculate the spin polarized conductance, it is relevant to study the fluctuations in $G^s_\alpha$ $(\alpha=x,y,z)$. Fig.\ref{dg} shows the fluctuations in the spin polarized conductance, $\Delta G^s_\alpha$ as a function of the Fermi energy. Though we did not find any similarity in the behaviour among the three components of the spin polarization, however the nature of fluctuations of the three components are exactly the same. The spikes in these fluctuation spectra can be understood if we look at Fig.\ref{spike}, where the variation of $\Delta G^s_y$ is shown for the ZGNR case in Fig.\ref{spike}(a) and for the AGNR case in Fig.\ref{spike}(c). In the lower panel, we plot the variation of the charge conductance, $G$ for the ZGNR case in Fig.\ref{spike}(b) and for the AGNR case in Fig.\ref{spike}(d) in the clean limit, that is, where the system is free from any kind of spin-orbit interaction. In this limit, $G$ shows a step-like behaviour. The height of the step depends on the number of modes available to the system for conduction for a given energy range. For each step-like feature in $G$, there is a peak in $\Delta G^s_y$. Thus we can say that, whenever a mode opens at a particular energy, the fluctuation in $\Delta G^s_y$ shoots up and thus there is a spike.    

Apart from the sign factor of the spin polarization, the qualitative behaviour of the ZGNR and AGNR setup are almost same. Hence, from now on, for brevity, we shall focus on the ZGNR setup only.

\begin{figure}[h]
\begin{center}
\includegraphics[width=0.5\textwidth]{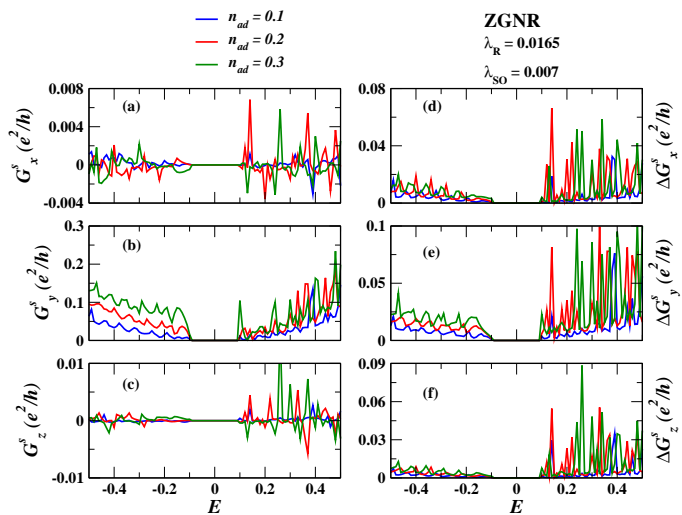}
\caption{(Color online) (a-c) All the three components of spin polarized conductance  and (d-f) their corresponding fluctuations are plotted as a function of Fermi energy for the ZGNR case only in case of Au adatom.}
\label{au-sp}
\end{center}
\end{figure}

The variation of spin polarized conductance and their fluctuations as a function of Fermi energy are shown in Fig.\ref{au-sp} in case of Au adatom. The $x$, $y$ and $z$-component of the spin polarization are plotted in Fig.\ref{au-sp}(a)-(c) and their corresponding fluctuations in Fig.\ref{au-sp}(d)-(f). Here owing to large fluctuation compared to the charge conductance case, 200 random-adatom configurations are taken. The magnitude of $G^s_x$ and $G^s_z$ are smaller by two orders of magnitude than $G^s_y$. Also $x$ and $z$ components of spin polarizations do not have any regular feature as a function of the Fermi energy. Most interesting point is that, when the Fermi energy exceeds a certain value, specifically for $E\sim0.1$, all the three components of spin polarizations and their fluctuations become substantially oscillatory in nature, which may be due to the following reason. The chemical potential for the Au adatom is taken as $\mu = 0.1$, so when the Fermi energy becomes equal or greater than the $\mu$ value, the screening of charges between the Au adatom and its neighbouring carbon atoms may not continue to occur. $G^s_y$  and $\Delta G^s_y$ increase as we increase the adatom density, which can be clearly noticed for $E<0$.

\begin{figure}[h]
\begin{center}
\includegraphics[width=0.5\textwidth]{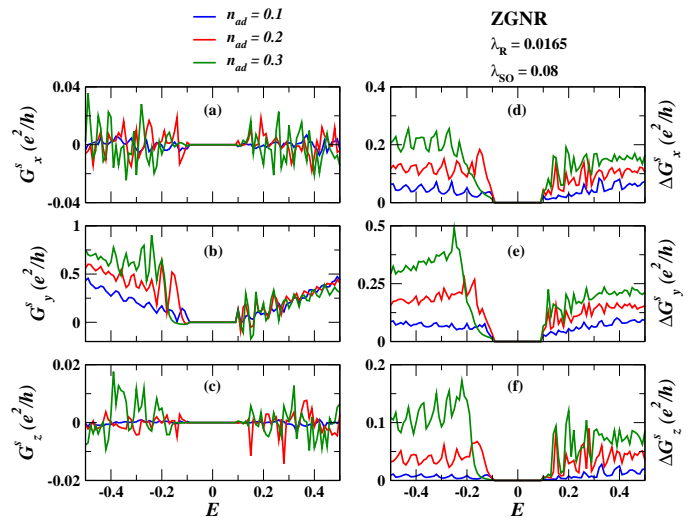}
\caption{(Color online) (a-c) All the three components of spin polarized conductance  and (d-f) their corresponding fluctuations are plotted as a function of Fermi energy for ZGNR case only with $\lambda_{SO}=0.08$ (hypothetical adatom). Rest of the parameters are same as Au adatom.}
\label{au_sp_lso0.08}
\end{center}
\end{figure}

By increasing the strength of intrinsic SOC, the results discussed above can be improved a little. In this case we set $\lambda_{SO}=0.08$. Rest of the parameters are taken exactly same as that of Au adatom. Fig.\ref{au_sp_lso0.08} shows the variation of $G^s_x$, $G^s_y$ and $G^s_z$ and their corresponding fluctuations as a function of Fermi energy for $n_{ad} = 0.1,0.2$ and $0.3$. The magnitudes of all the three components of spin polarization increase than the case corresponding to Au adatoms (Fig.\ref{au-sp}), along with reduced oscillatory nature for $E\geq 0.1$. Fig.\ref{au_sp_lso0.08}(d-f) show the behaviour of $\Delta G^s_x$, $\Delta G^s_y$ and $\Delta G^s_z$. Where they show very similar behaviour as a function of the Fermi energy. One can ask why the fluctuations have greater magnitude than their observables. Since the spin polarized conductance can have both positive as well as negative values and also since their magnitudes are close to zero, there is a finite possibility that the averaged value can be close to zero. On the other hand, if we recall the expression for the fluctuations (see Eq.\ref{fl}), the term $\langle \left(G^s_\alpha\right)^2\rangle$ is always is greater than zero and additive in nature. As a result, $\Delta G^s_\alpha$ assumes large (and oscillatory) values. 
%--------------------------CONCLUSION-----------------------------------------------------
\section{\label{sec6}Conclusion}
In the present work we have studied the behaviour of charge and spin polarized conductances in zigzag and armchair graphene nanoribbons with Au adsorbates. Au as adatom is chosen since it induces Rashba SOC in the ribbon. We made a comparison between ZGNR and AGNR and if the width of AGNR is taken as $N_y=3M-1$ (M being an integer) which makes AGNR metallic, both type of ribbons show more or less same qualitative behaviour. In case of Au adatom, the systems are no longer in QSH  phase. However if we enhance the strength of the intrinsic SOC, the QSH effect  can be restored. By introducing the adatom in the system, we effectively break the longitudinal mirror symmetry and hence we are able to generate the $x$ and $z$-component of the spin polarizations. However their magnitude is one or two orders smaller than than that of the $y$-component of the spin polarized conductance. The most interesting point is that, the fluctuations of all the components of spin polarization in presence of Rashba SOC only, have the same behavior, which however goes away in case of Au adatoms.

It may be noted that we have considered an artificial adatom which can induce an intrinsic SOC at least one order larger than that is possible for Au adatoms. Thus careful scrutiny of different (heavy) elements of the periodic table that can induce larger SOC in graphene nanoribbons is needed.
\\

\setcounter{secnumdepth}{0}
\section{ACKNOWLEDGMENTS}
SG gratefully acknowledges a research fellowship from MHRD, Govt. of India.
SB thanks SERB, India for financial support under the grant F. No: EMR/2015/001039.

\end{document}